\documentclass[]{spie}

\usepackage{amsmath,amsfonts,amssymb}
\usepackage{graphicx}
\usepackage[colorlinks=true, allcolors=blue]{hyperref}
\usepackage{subfig}
\usepackage{enumitem}

\title{MEAD: Data Reduction Pipeline for ALES Integral Field Spectrograph and LBTI Thermal Infrared Calibration Unit}

\author[a]{Zackery Briesemeister}
\affil[a]{Department of Astronomy and Astrophysics, University of California, Santa Cruz, CA 95064}
\authorinfo{zbriesem@ucsc.edu}

\author[a]{Andrew J. Skemer}

\author[b]{Jordan M. Stone}
\affil[b]{Steward Observatory, University of Arizona, Tuscon, AZ 85721}

\author[a]{R. Deno Stelter}

\author[b]{Philip Hinz}

\author[b]{Jarron Leisenring}

\author[c]{Michael F. Skrutskie}
\affil[c]{Department of Astronomy, University of Virginia, Charlottesville, VA 22904}

\author[d]{Charles E. Woodward}
\affil[d]{Minnesota Institute for Astrophysics, University of Minnesota, Minneapolis, MN 55057}

\author[e]{Travis Barman}
\affil[e]{Lunar and Planetary Laboratory, University of Arizona, Tucson, AZ 85721}

\begin{document} 
\maketitle

\begin{abstract}

We present the data reduction pipeline, \texttt{MEAD}, for Arizona Lenslets for Exoplanet Spectroscopy (ALES), the first thermal infrared integral field spectrograph designed for high-contrast imaging. ALES is an upgrade of LMIRCam, the $1-5\,\mu$m imaging camera for the Large Binocular Telescope, capable of observing astronomical objects in the thermal infrared ($3-5\,\mu$m) to produce simultaneous spatial and spectral data cubes. The pipeline is currently designed to perform $L$-band ($2.8-4.2\,\mu$m) data cube reconstruction, relying on methods used extensively by current near-infrared integral field spectrographs. ALES data cube reconstruction on each spectra uses an optimal extraction method. The calibration unit comprises a thermal infrared source, a monochromator and an optical diffuser designed to inject specific wavelengths of light into LBTI to evenly illuminate the pupil plane and ALES lenslet array with monochromatic light. Not only does the calibration unit facilitate wavelength calibration for ALES and LBTI, but it also provides images of monochromatic point spread functions (PSFs). A linear combination of these monochromatic PSFs can be optimized to fit each spectrum in the least-square sense via $\chi^2$ fitting. 

\end{abstract}

\section{INTRODUCTION}
\label{sec:intro} 

Arizona Lenslets for Exoplanet Spectroscopy (ALES \cite{2015SPIE.9605E..1DS}) is a project designed to extend the functionality of the Large Binocular Telescope Interferometer's (LBTI \cite{2008SPIE.7013E..39H, 2008SPIE.7013E..28H, 2012SPIE.8445E..0UH, 2014SPIE.9146E..0TH}) $1-5\,\mu$m imager LMIRCam \cite{2010SPIE.7735E..3HS,2012SPIE.8446E..4FL}. ALES is the first integral field spectrograph (IFS) capable of high-contrast imaging in the thermal infrared. A current scientific goal utilizing ALES on a single aperture of LBT is to deliver low-resolution $LM$-band spectra of young, gas giant exoplanets and substellar companions in order to supplement existing near-infrared $JHK$ spectra for a broader spectroscopic characterization of these bodies. This goal can be accomplished by exploiting the unique properties of IFS data cubes, which comprise photometrically accurate stacks of simultaneous narrowband images spanning multiple wavelength channels. The spatial and spectral information within the IFS cubes enables unambiguous separation of the light from substellar companions and their host star.

The success of near-infrared lenslet-based IFSs has been due in large part to the development of robust techniques for automating the construction of wavelength-calibrated spectral data cubes from the thousands of closely-packed spectra in raw frames (e.g. GPI \cite{2008SPIE.7015E..18M, 2014SPIE.9147E..3JP} ; SPHERE \cite{2008SPIE.7014E..3EC, 2008SPIE.7019E..39P} ; Project 1640 \cite{2011PASP..123...74H, 2011PASP..123..746Z} ; CHARIS \cite{2012SPIE.8446E..9CM, 2017JATIS...3d8002B} ; OSIRIS \cite{2006SPIE.6269E..1AL}). These pipelines are critical for the homogenization of data products and the accessibility for other observers.

The Methods for Extracting ALES Data (\texttt{MEAD}) package is the Python-language data reduction pipeline for ALES that has leveraged the insights gained during the operations of the near-infrared IFSs in order to orchestrate the construction of ALES data cubes. After a brief overview of the instrument, this paper begins by presenting \texttt{MEAD} in a linear fashion, following a recipe with which most observations will be reduced. Then the paper focuses on the thermal infrared calibration unit for LBTI and how the unit will affect ALES operations. Section \ref{sec:det} summarizes basic processing of raw data frames to remove detector artifacts. Section \ref{sec:fpm} addresses the extent and characterization of flexure in the instrument, as well as the calibration process. Section \ref{sec:cube} briefly states how the cubes are reconstructed. Section \ref{sec:ticu} cover the thermal infrared calibration unit for LBTI. We finish by discussing the immediate future for ALES and \texttt{MEAD} in Section \ref{sec:future}.

\section{Instrument Overview}

ALES\cite{2015SPIE.9605E..1DS} is a low resolution, thermal infrared, lenslet-based integral field spectrograph inside LBTI \cite{2008SPIE.7013E..39H, 2008SPIE.7013E..28H, 2012SPIE.8445E..0UH, 2014SPIE.9146E..0TH}/LMIRCam \cite{2010SPIE.7735E..3HS,2012SPIE.8446E..4FL}. Unlike the near-infrared analogues, LMIRCam is not solely dedicated to science performed with ALES; the ALES optics are set inside filter wheels in order to allow LMIRCam operations to remain undisturbed. ALES itself comprises an 8$\times$ Keplerian magnifier, a silicon lenslet array, a diffraction suppressing pinhole grid, direct-vision prism and blocking filter inside LBTI/LMIRCam. 

LBTI is situated at the combined bent Gregorian focus of the LBT's twin 8.4 meter mirrors, each equipped with deformable secondary mirrors \cite{2011SPIE.8149E..02E}, LBTI adaptive optics (AO) systems \cite{2014SPIE.9148E..03B}, and flat tertiaries that redirect wavefront-corrected, diffraction-limited, infrared light into the cryogenic universal beam combiner (UBC). The initial mode of ALES was designed to operate behind a single aperture of LBT, so the UBC redirects the light without any pathlength correction into the science instrument dewar (Nulling Infrared Camera, NIC) and the cryogenic science camera, LMIRcam. 

At the focal plane of each lenslet, an image of the exit sub-pupil will form, comprising of all the light from the image incident on the spatial extent of the lenslet. The subpupils are then dispersed without spatial or spectral overlap, forming images of dispersed subpupils on LMIRCam's Teledyne HAWAII-2RG (H2RG \cite{2008SPIE.7021E..0HB}). Each subpupil becomes a spatial pixel element (spaxel) with an associated spectrum in the final spectral data cubes. At the focal plane, the light is dispersed at a fiducial angle of $\theta =tan^{-1}\frac{1}{2} \approx 26.56^{\circ}$ with respect to the detector columns. For its inital mode for which this paper is most relevant, ALES observes $L$-band ($2.8-4.2 \,\mu$m) spectral data cubes. In detector coordinates, the $L$-band spectra are expected extend 35.6 pixels in the spectral direction and 4-5 pixels FWHM in the spatial direction \cite{2015SPIE.9605E..1DS}.

ALES first light, design, data processing, operations, and upcoming upgrades are discussed in the following:
\begin{itemize}[noitemsep]
\item First Light with ALES: A 2-5 Micron Adaptive Optics Integral Field Spectrograph for the LBT (Skemer et al. 2015 \cite{2015SPIE.9605E..1DS})
\item Design of ALES: a broad wavelength integral field unit for LBTI/LMIRCam (Hinz et al., paper \#10702-130)
\item MEAD: Data Reduction Pipeline for ALES Integral Field Spectrograph and LBTI Thermal Infrared Calibration Unit (This work)
\item On-sky operations with the ALES integral field spectrograph (Stone et al., paper \#10702-124)
\item ALES: Overview and Upgrades (Skemer et al., paper \#10702-11)
\end{itemize}

Looking towards the future of ALES observations, we also encourage readers to consult (Spalding et al., paper \#10701-68) and (Leisenring et al. 2014 \cite{2014SPIE.9146E..2SL}) for a discussion of Fizeau interferometry using dual aperture mode of LBT with LBTI. This will allow ALES to observe at the diffraction limit with the entire 23.4m effective aperture of the LBT.

\section{Data Reduction Overview}

While ALES has unique science capabilities, ALES science can only be as good as its data reduction pipeline. As operations have matured, \texttt{MEAD} has been developed to approach the problem of extracting the science data from raw frames with the goal of user-friendliness and flexibility. While the main focus of the pipeline is to produce photometrically accurate and calibrated data cubes for scientific analysis, we also require quick reductions to guide ALES operations. This will also include guiding on the occulted point spread function from one of LMIRCam's coronagraphs, following its recent alignment with ALES optics. 

Ultimately, \texttt{MEAD} will exist as a software package to cover all stages of data reduction and post-processing for ALES. However, the complete automation and the integration with LBTI/LMIRCam software remains to be implemented. \texttt{MEAD} has endured rapid development in order to address the expected updates to ALES for 2018B (Skemer et al., paper \#10702-11) and will soon be critical to controlling ALES operations.

The full science data reduction is organized into five main components as follows:

\textbf{Data Parser} -- The data parsing tool is designed to reconstruct the progression of an observation run from metadata and organize a data set into their associated sequences. This can be performed na\"ively with just the FITS header information in each science exposure. However, ancillary information regarding the pointing pattern and well-documented observing logs are necessary to intelligently organize the data and exclude bad frames, respectively. Changes to the pointing pattern should be updated manually in the configuration files.

\textbf{Basic Processing} -- User-discretion methods for bias subtraction, dark subtraction, pixel flat-fielding, frame combination, bad pixel identification and removal, linearity correction, and microphonic noise suppression. The merits and downsides of certain corrections are addressed in Section \ref{sec:det}.

\textbf{Focal Plane Geometry Calculation} -- ALES is a filter wheel instrument, so the lenslet array, dispersing element (prism) and detector are not absolutely static with respect to one another; mechanical flexure that distorts the focal plane geometry manifests with strong field dependence. In order to extract each spectrum as homogeneously as possible, this stage develops of a piecewise focal plane model, in which each spectrum is calibrated independently. Every light-sensitive pixel is then mapped to a weight in a wavelength-calibrated data cube. See Section \ref{sec:fpm} for further explanation.

\textbf{Cube Construction} -- \texttt{MEAD} orchestrates the construction of data cubes. For optimal or aperture extraction methods, the focal plane solution is used to map every light-sensitive pixel to a weight in a data cube, and the flux extraction method defines the magnitude of these weights. See Section \ref{sec:cube} for further explanation. For the $\chi^2$ extraction method described in Section \ref{sec:chi}, cubes can be built from the linear combination of position- and wavelength-dependent monochromatic point spread functions that fit each spectrum. 

\textbf{Post-Processing} -- For high-contrast imaging, PSF subtraction using Angular Differential Imaging (KLIP \cite{2012ApJ...755L..28S}) is currently implemented. We will expand to other PSF subtraction methods when relevant. ALES-specific wrappers for the Vortex Image Processing (VIP \cite{1538-3881-154-1-7}) package implementation of other post-processing methods are also available.

Quicklook reductions require far less robust calculation, especially because the coronagraph can effectively be guided with any individual wavelength channel image or even the wavelength-collapsed cube. The relative aggressiveness of reduction can be controlled manually, but are also associated with longer runtime. The optimization of realtime reduction and post-processing remains an open investigation.

\section{Basic Processing} \label{sec:det}

The science detector for ALES is the 2048$\times$2048 pixel H2RG of LMIRCam. Early observations were performed with FORCAST \cite{2010SPIE.7735E..2NL} readout electronics, which limited the detector readout to a 1024$\times$1024 pixel sub-array. This sub-array has 50$\times$50 spatial pixel elements (spaxels) with a field of view of 1.3"$\times$1.3". The readout electronics have since been replaced with Teledyne SIDECAR ASIC \cite{2005SPIE.5904..293L} electronics to read out the entire detector. This section will give an overview of corrections for detector artifacts performed during a standard reduction.
\subsection{Detector Artifacts}
\subsubsection{Residual Channel Bias}

Each of the 64-pixel-wide channels has its own analog-to-digital converter set an unique bias levels. This poses a problem for spaxels that overlap two readout channels, particularly because the offset would propagate to the reduced cube as a striping pattern that affects different spaxels at different wavelengths.

The residual channel bias offsets are corrected using the median of each channel's reference pixels, which are then subtracted off from their respective channels. Residual channel bias is not observed to be constant or repeatable in successive frames, so this correction must occur after dark- or thermal background-subtraction.

\subsubsection{Flatfielding}
Integral field spectrographs have two types of flatfielding: pixel and lenslet flatfields. The traditional pixel flatfielding adjusts for variable detector gain, while lenslet flats are a characterization of variable lenslet throughput as a function of wavelength. Pixel flatfields are a correction in detector coordinates, while lenslet flatfields are a correction in spaxel coordinates. Lenslet flatfields will be discussed in Section \ref{sec:cube}.

Without intervention, the pipeline defaults to building and using a pixel flatfield if the data parser flags such data to exist. This functionality can be toggled in the configuration file. The thermal background (thermal telescope and sky emission) is imaged without ALES optics on LMIRCam, which are then dark-subtracted, median-combined, and divided by the mode to create a master flat. Pixels with a correction greater than 1.5 or less than .5 are flagged to be bad pixels.

\subsubsection{Dark Subtraction}
Median dark frames are subtracted from thermal background frames for cube reconstruction of the lenslet flatfield (See Section \ref{sec:cube}). Narrowband calibration data must also be dark-subtracted in order to remove detector artifacts. In both of these cases, the residual channel bias correction is necessary.

\subsubsection{Bad Pixel Correction}

\texttt{MEAD} comes with a reference bad pixel mask, which can be updated, remade, or completely disregarded. The reference bad pixel mask was made from LMIRCam data from 2016 with the same method in which the pipeline would remake the mask upon request. We calculated the standard deviation of a stack of dedicated dark frames. Since the distribution was asymmetric, we chose to flag the top 2\% and bottom 1\% as bad pixels. If pixels were flagged during pixel flatfield creation, they are also flagged here. Bad pixels are corrected by replacing them with the average of the nearest four good neighboring pixels. Cosmic rays are not a significant noise source, as the H2RG is substrate-removed. Existing cosmic rays are nominally removed when median combining frames.

\subsubsection{Linearity}

\texttt{MEAD} also comes with reference linearity correction, which can be used, remade or ignored. Ten dark-subtracted flatfield frames are imaged at each exposure time possible between .03 and 2.33 seconds. For a uniform correction across the entire array, a linear fit to the linear part of the medians of the ten frames for each exposure time is used to define the linearity correction. For the field-dependent correction, a linear fit to every pixel in the time-series frames defines the linearity correction. The correction is set to only apply near where the fit begins to diverge .1\% from the observed linearity data. 

\subsubsection{Microphonic Noise Suppression}

Microphonic noise induced by vibrations from an undetermined source manifest as horizontal sinusoidal patterns on the detector. Although microphonic noise is only observed for short exposure times, the thermal background limits going to longer exposure times. We remove the microphonics artifact in a similar fashion to GPI by diminishing the intensity of frequencies corresponding to the noise in the image proportionately to the dot product of the image with a noise model built from short exposure time dark frames \cite{2010SPIE.7735E..31M, 2014SPIE.9147E..3JP}.

\section{Focal Plane Geometry} \label{sec:fpm}

The initial mode of ALES is designed to deliver to the focal plane 2,500 equally-spaced $2.8-4.2 \,\mu$m spectra, each spanning 35.6 pixels in the dispersion direction at an angle $\theta = tan^{-1}\frac{1}{2} \approx 26.56^{\circ}$. Mechanical flexure distorts the geometry of the spectrograph focal plane, changing the position and dispersion angle of science spectra with respect to calibration spectra. The deviation from the fiducial dispersion angle, the shift along the dispersion axis and the shift perpendicular to the dispersion axis are the three parameters necessary to describe and calibrate the distortion from flexure. These three parameters for each spectrum also have field dependence (Figure \ref{fig:fielddep}). This requires a joint characterization of these deviations in order to have calibrate the spectra in the cube.

\begin{figure}[h]

\centering
\includegraphics[trim=4cm 0cm 0cm 0cm, scale=.4]{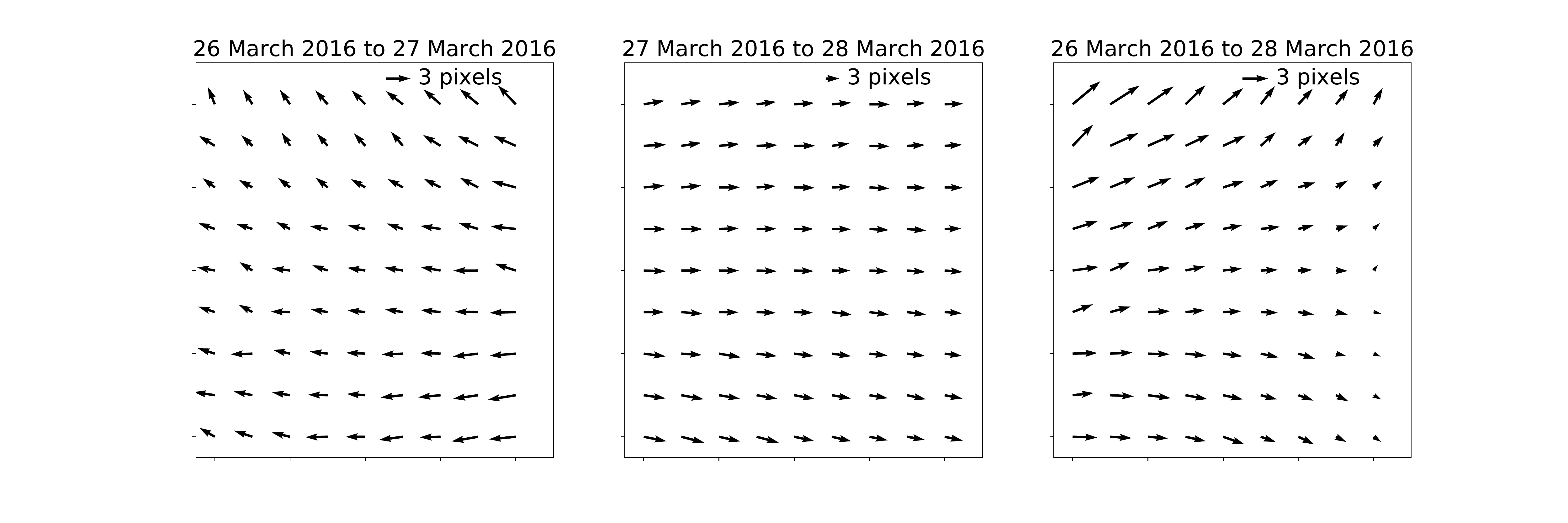}
\caption{The evolution of the spectrograph focal plane geometry over a three night period. Each arrow represents the median deviation between the spaxel position of two nights in a 5$\times$5 spaxel region. The spaxel position is defined by the position of the 3.950 $\mu$m narrowband filter spot associated with the spaxel. 
}
\label{fig:quiver}
\end{figure}

The focal plane geometry differed on a night-to-night basis by almost 3 pixels because the alignment of the ALES optics within filter wheels is not perfectly reproduced on this timescale (Figure \ref{fig:quiver}). Instead of favoring some smooth distortion to calibration data, we prefer to simply take a complete set of calibration data for every night.

\subsection{Crosstalk}\label{ssec:cross}

ALES is designed such that the separation of the spectra is maximized, which minimizes the crosstalk between adjacent spaxels. However, astigmatism induced by the second biconic mirror inside LMIRCam manifest in ALES raw data as a decay in the sharpness of spatial profiles near the edge of the detector, and these spectra have nontrivial crosstalk with one another. As part of ALES upgrades in 2018, the lenslet array is being replaced with a lenslet array that reverses the astigmatism induced by the mirror and should successfully suppress the crosstalk effect near the edges. Apart from using more appropriate weights in the optimal extraction (Section \ref{sec:cube}), we will wait for the new lenslet array upgrade prior to developing a more robust correction.

The $\chi^2$ extraction method described in Section \ref{sec:chi} is ideal for addressing crosstalk, as the monochromatic PSFs would also experience the same optical distortions that cause the spatial and spectral resolution to decay across the field. The spectra are first fit na\"ively, and the resulting model 2D spectra are subtracted from the data. The individual model spectra overlap with neighboring spectra, resulting in negative residuals. An iterative approach to minimizing the magnitude of the negative residuals is then applied to remove the effects of crosstalk.

\subsection{Isolating the Spectra}

ALES has access to four narrowband filters with central wavelengths 2.925, 3.375, 3.555, 3.950 $\mu m$ for $L$-band calibrations. ALES observes the thermal background with the narrowband filters. Since the thermal background is almost two orders of magnitude brighter in the red end of $L$-band compared to the blue end, the 3.950 $\mu m$ is the brightest narrowband filter imaged with ALES. Each spectrum is coarsely located using the pixel location of the peak of the associated 3.950 $\mu m$ narrowband filter data. A $60\times40$ pixel rectangular slice is defined around each spectrum, with enough room such than the extraction region for the spectrum would remain populated with data given shifts parallel and perpendicular to the dispersion axis. The small rectangular slice reduces computational cost of interpolation.

The 3.950 $\mu m$ narrowband filter data peak is located by centroiding, which would, in principle, give an origin with which to rotate the extraction region. However, for calibration data taken at a different gravity vector than the target pointings, the narrowband filter data might not be coincident with the science data. 

\subsection{Solution for Offsets} \label{ssec:off}

We have since adjusted the observing pattern for ALES to include calibrations taken without slewing the telescope away from the science target. Under the same gravity vector, the perpendicular and parallel shifts with respect to the dispersion axis are minimized to subpixel magnitudes. However, calculating deviations from the fiducial dispersion angle is still required for the rectification process.

The deviation from the fiducial dispersion angle, along with perpendicular and parallel shifts along the dispersion axis from the calibration data become three offset parameters for every spectrum. The flexure between target pointings and sky pointings were empirically determined to be negligible under normal observing conditions, so the even illumination of the thermal background during sky pointings is ideal for calibration.

The sky spectrum in $L$-band, modulated by filter transmission, has features that can be used to calculate parallel shifts. Cross-correlation between the ALES sky spectral cubes with an ALES-resolution model spectra of the thermal background allows for wavelength shifts to be calibrated. Perpendicular shifts are derived by differencing the centroids of the spatial profiles of the sky spectra and all narrowband filter data.

\begin{figure}[h]
\begin{tabular}{ll}
\includegraphics[trim=2cm 1cm 0cm 1cm, clip=True, scale=.6]{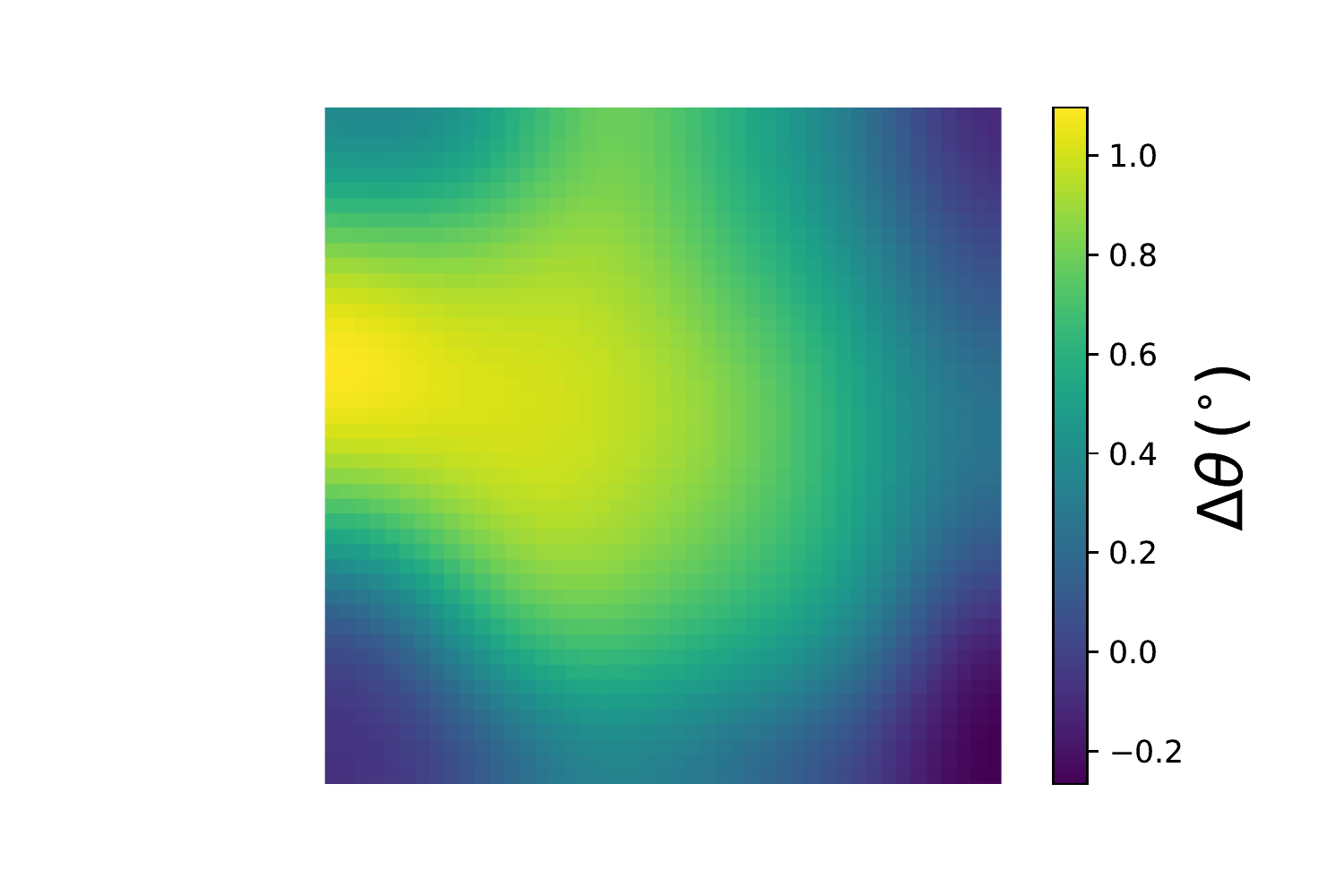}
&
\includegraphics[trim=2cm 1cm 0cm 1cm, clip=True, scale=.6]{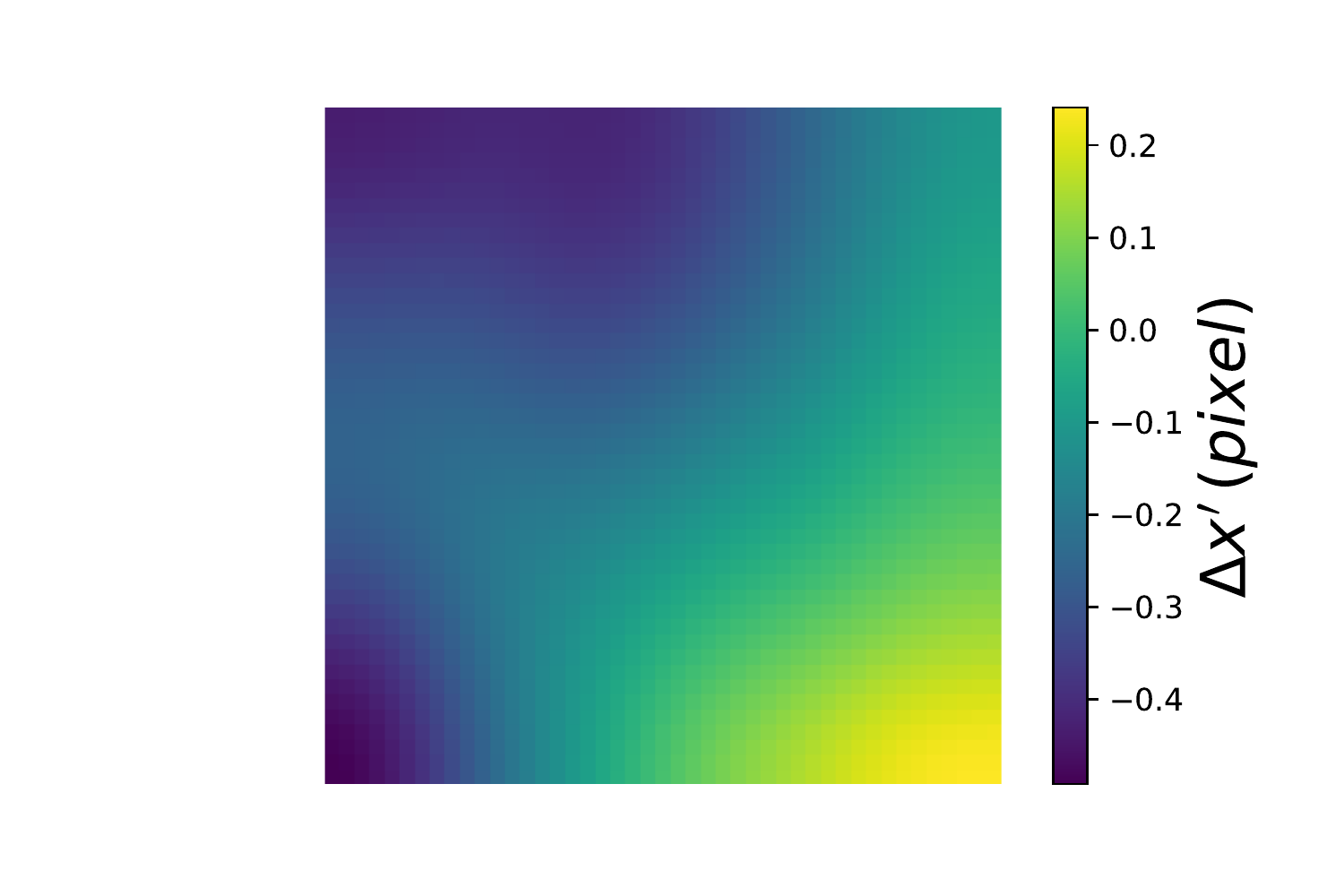}
\end{tabular}
\caption{Left: The field dependent deviation of the dispersion angle from $\theta = tan^{-1}\frac{1}{2}$. The average angle excess on 27 March 2016 is $0.55^{\circ}$. 
Right: The field dependent perpendicular shift from the dispersion axis. The average spatial offset of calibration data from science data on 27 March 2016 was -0.17 pixels. Not shown: the field dependent parallel shift along the dispersion axis of the calibration data from the science data. The magnitude was determined to be consistent with no shift.}
\label{fig:fielddep}
\end{figure}

The deviation from the dispersion angle is calculated using a Radon transform, performed via a grid search of rotation angles to populate a sinogram with the spatial profiles of rotated, mean-subtracted 2D spectra. The dispersion angle can be identified using one of two methods: the angle that minimizes the root-mean-square of the mean-subtracted average spatial profile or the angle that minimizes the absolute magnitude of the slope of the line of best fit of the centroids of the spatial profiles. The resulting field dependent deviations are then Gaussian smoothed with FWHM of one spaxel (Figure \ref{fig:fielddep}).

The wavelength solution is built from fitting an empirical dispersion relation to the peaks of the narrowband filter data. By default, the pipeline chooses the wavelength solution of the central spaxel to be the wavelength solution for the final data cube. The solver works by updating the coordinates for an affine transformation; after all the calibration, only one interpolation per spectrum will be necessary for both flux extraction and wavelength calibrate on a common wavelength grid such that the data cube can be easily visualized and manipulated. The single interpolation minimizes distortion of the data and homogenizes the cube construction across all spectra.

The pipeline can build as many calibration data files as there are thermal background realizations. However, there is not significant flexure between pointing patterns, and it is enough to calibrate on one or two thermal background observations for a whole night. 

\section{Cube Construction} \label{sec:cube}

With the data products from the focal plane geometry calculation, \texttt{MEAD} constructs the data cubes from the raw frames using the same coordinate transformations derived during the calibration step. This cube construction, as well as the calibration step, act on subsections of the entire frame completely independent from one another. This makes these steps ideal for parallelization, and they have been implemented as such. The user can also toggle the parallelization and number of processors in the configuration file.

The current preferred method for performing spectral extraction for cube construction is optimal extraction \cite{1986PASP...98..609H}. This method combines empirical information about the spatial profiles and errors to provide informed weights for a weighted-average. The field dependent astigmatism causes spatial profiles to vary across an ALES field that would otherwise be ignored under a unity-weighted scheme, and optimal extraction is not computationally more expensive. The optimal extraction method uses the empirical spatial profiles derived from the thermal background frames.

\subsection{Data Products}

\begin{figure}[h]
\includegraphics[trim=5cm 2.5cm 0cm 3.5cm, clip=True, scale=.41]{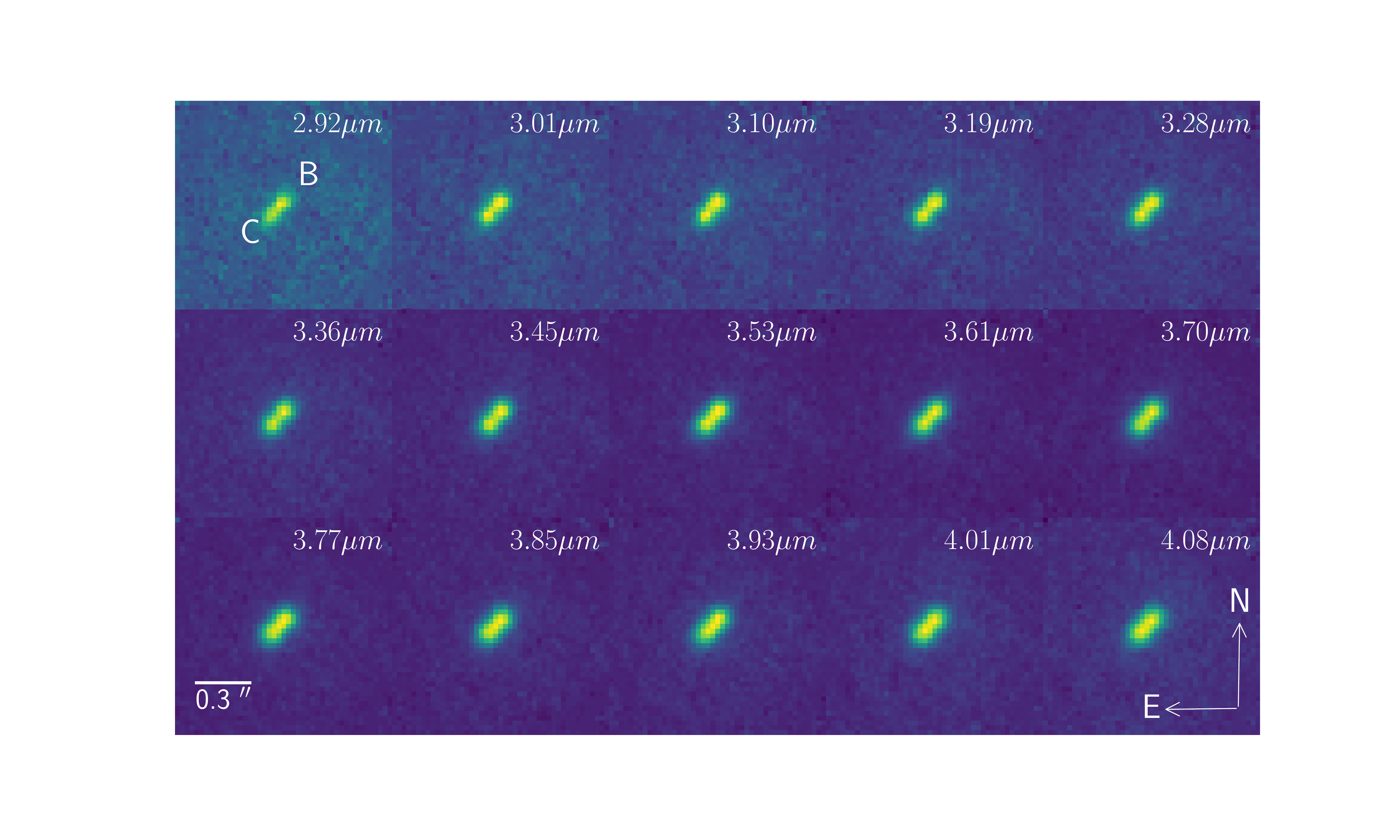}

\caption{An example of an ALES data cube, with 15 representative wavelength slices of the data cube for HD 130948 BC, a L4+L4 benchmark brown dwarf binary orbiting as sun-like primary star in a hierarchical system. The pair of brown dwarfs are separated by .104", and are resolved in $L$-band with ALES.
}
\label{fig:jordanplot}
\end{figure}

The analyzable data products include science target data cubes, spectrophotometric calibration data cubes, astrometric calibration data cubes, lenslet flatfields and the wavelength solution for the data cubes. All data cubes are also accompanied by their variance data cube, which have been propagated from the variance image of raw frames. Slices of an example science target data cube can be seen in Figure \ref{fig:jordanplot}.

The lenslet flatfield is built from constructing cubes from dark-subtracted thermal background frames that are then normalized at each wavelength channel by the median of the slice. The lenslet flatfield is then a data cube with each wavelength channel being the field dependent throughput of each lenslet. Typical values in the lenslet flatfield are between .95 and 1.05, with extrema at .90 and 1.10.

For high-contrast imaging datasets, these data products are prepared for post-processing. \texttt{MEAD} contains limited post-processing tools in its current state, namely Karhunen-Loeve Image Projection (KLIP \cite{2012ApJ...755L..28S}) for Angular Differential Imaging. The science goals of ALES are currently guiding the implementation of more post-processing tools. \texttt{MEAD} also contains ALES-specific wrappers for the Vortex Image Processing (VIP \cite{1538-3881-154-1-7}) package implementation of Locally optimized Combination of Images (LOCI \cite{2007ApJ...660..770L}) and LLSG \cite{2016A&A...589A..54G}, but has not been extensively tested with all of the available functionality of the package. 

\section{Thermal Infrared Calibration Unit} \label{sec:ticu}

Robust calibration in the thermal infrared is tricky because the highly-variable and bright thermal background. Internally, LMIRCam contains the four narrowband filters, which are used to image narrowband flats using their transmission of the thermal background. This is sufficient for constraining the wavelength solution for the optimal extraction method, but the four narrowband filters do not provide a full coverage of monochromatic PSFs necessary to perform the $\chi^2$ extraction. External calibration sources would need to produce monochromatic flats that are both bright and stable with respect to the thermal background. The calibration unit also needs to be tunable in order to image the position- and wavelength-dependent monochromatic PSFs. 

\begin{figure}[h]

\centering
\includegraphics[trim= 1.8cm 8.75cm 0cm 8.75cm, scale=.95, clip=False]{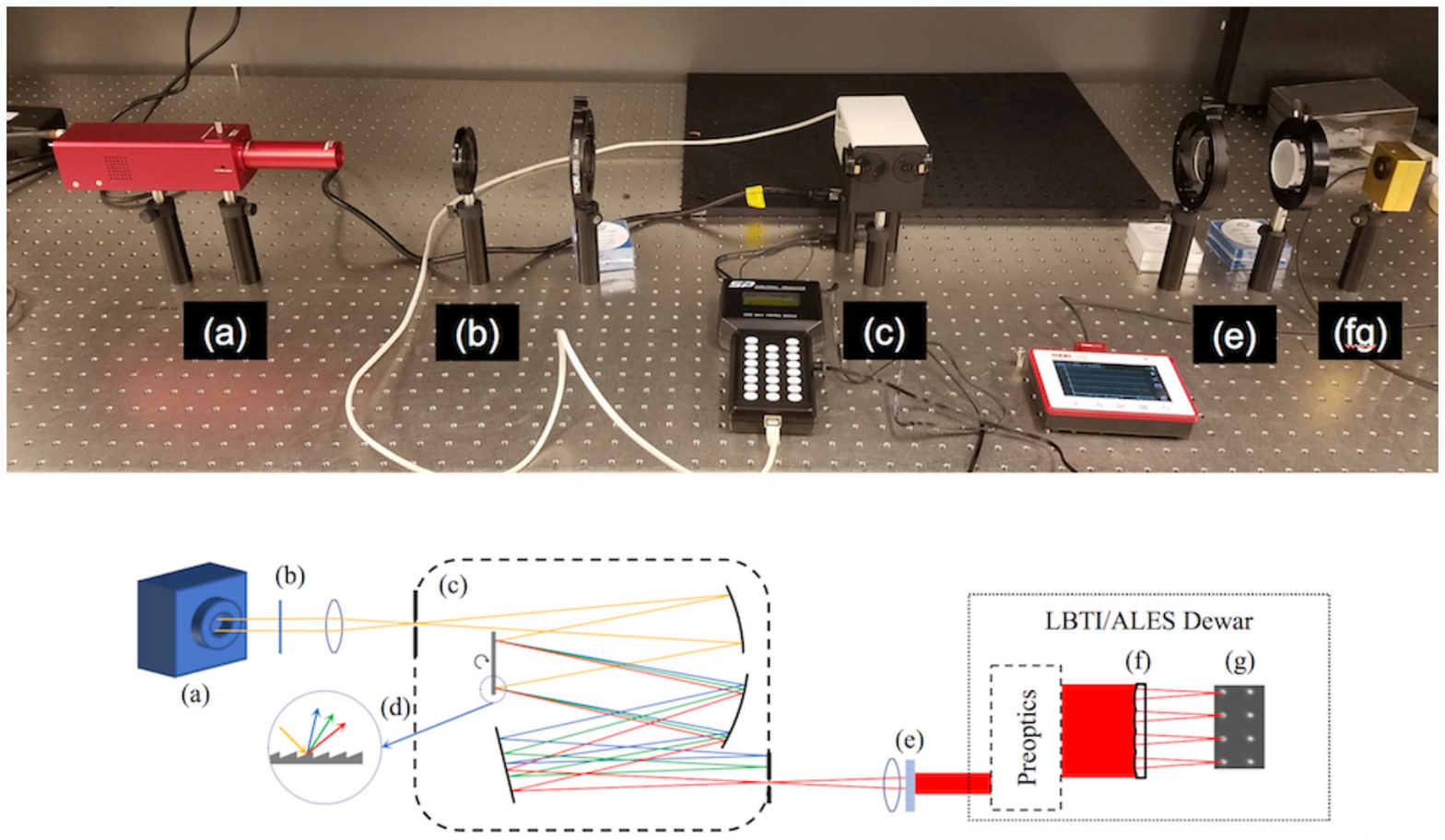}
\caption{Schematic and Lab image of the thermal infrared calibration unit for LBTI. The unit comprises the thermal infrared source (a), shutters and baffling (b; not depicted in image), a Czerny-Turner monochromator (cd) and the optical diffuser (e). An integrating sphere (fg) substitutes for LBTI in the lab.}
\label{fig:calunit}
\end{figure}

A new calibration method has been designed for LMIRCam/ALES with the goal to select wavelengths from a blackbody continuum to image monochromatic light on the detector. The calibration unit (Figure \ref{fig:calunit}) comprises a thermal infrared source (a), a Czerny-Turner monochromator (b) and an optical diffuser (e). The light from the thermal infrared source is focused onto the entrance slit of the monochromator, which has an rotatable grating (d) that disperses the light onto an exit slit. A specific wavelength of light is selected by rotating the grating, which translates the spectrum across the exit slit. This monochromatic light passes through an optical diffuser in order to evenly illuminate the pupil plane of ALES (fg). 

The design of the thermal infrared calibration unit (Figure \ref{fig:calunit}) is distinct from near-infrared implementations for imaging monochromatic PSFs because common optical materials are opaque at these wavelengths and thermal infrared photons increase the temperature of elements in the unit. Optics closer to the thermal infrared source are more likely to be heated and contribute to the background, requiring calibrations to be taken at steady state. This steady state may not be reachable in the time allotted for calibrations, requiring  background frames to be taken more often. 

The monochromator has two gratings designed for 400-1500 nm and 1500-6000 nm, respectively. A HeNe laser (632.8 nm) was used to confirm the wavelength calibration and design specifications of the first grating. However, there remains no independent confirmation of the wavelength calibration of the relevant 1500-6000 nm grating, but manufacturer supplied data suggests good conformity. Thermal infrared sources do not stop emitting thermal infrared photons when they are turned off for obtaining background frames in between calibrations frames; a system of shutters is necessary to block the light from entering LBTI as well as block the heat from previous shutters that will begin to glow from the incident radiation. Baffling and the shutter system (b) is not accurately represented in the lab setup in Figure \ref{fig:calunit} in order to show the rest of the unit.

It would be particularly difficult to illuminate the dome or a screen with monochromatic thermal photons in order to fill the primary with monochromatic light, so this calibration light will not be experiencing aberration from the three warm optics associated with each aperture. Instead, the calibration unit light will be incident on the entrance window of LBTI/NIC, and will still experience the astigmatism caused by the biconic mirror inside LMIRCam. It is key to determine the frequency of which a full coverage of monochromatic PSF data would need to be taken.

This thermal infrared calibration unit was built and tested at University of California, Santa Cruz, but is not on the Large Binocular Telescope yet.

\section{Future Work} \label{sec:future}
The data reduction pipeline is also accommodating the multitude of upgrades to the instrument coming in 2018B. Apart from the thermal infrared calibration unit, the extent of these upgrades are described in (Skemer et al., paper \#10702-11).

\subsection{Chi Square Extraction} \label{sec:chi}

The least-square inversion flux ($\chi^2$) extraction method \cite{2017JATIS...3d8002B, 2014SPIE.9147E..4ZD} performs spectral deconvolution by proposing a linear combination of empirical monochromatic PSFs to fit each spectrum in the least-square sense. This method would be particularly powerful to implement for ALES because the field dependent astigmatism distorts the monochromatic PSFs in the same way it distorts the spectra.

The method is advantageous because it circumvents artifacts associated with the optimal extraction method: it performs no interpolations that would induce spectral correlation and it uncouples the spectrum from systematic noise. It trivially fits the undispersed background, and avoids final interpolation onto a common wavelength grid, which is a source of spectral correlation induced by current methods of flux extraction.

The thermal calibration unit facilitates imaging a full coverage of the position- and wavelength-dependent monochromatic point spread functions that would make the $\chi^2$ extraction method possible for ALES. The implementation currently exists using singular value decomposition in \texttt{MEAD} and has been tested on simulated data, but extensive testing with real data from the calibration unit will be required before ALES operations use this method of extraction.

\subsection{Coronagraph Alignment and Guiding}

ALES operations have as of yet not included the use of a coronagraph, which will eventually become key to probing higher contrasts. ALES is aligned with LBTI's $L^{\prime}$ annular groove phase mask coronagraph \cite{2014SPIE.9148E..3XD} inside LMIRCam's filter wheel, but guiding on an obstructed PSF from raw integral field spectrograph frames presents an obstacle that will need to be overcome for 2018B. The extent at which the full data reduction procedure can be pared down in favor of expediency and accuracy for guiding the coronagraph will need to be tested, largely dependent on the magnitude of flexure. This quick-look version of data reduction would forgo rigorous calibration, and simply takes in narrowband filter data or uses an old focal plane model. This remains to be integrated with LBTI/LMIRCam software.

\acknowledgements
The LBT is an international collaboration among institutions in the United States, Italy and Germany. LBT Corporation partners are: The University of Arizona on behalf of the Arizona university system; Istituto Nazionale di Astrofisica, Italy; LBT Beteiligungsgesellschaft, Germany, representing the Max-Planck Society, the Astrophysical Institute Potsdam, and Heidelberg University; The Ohio State University, and The Research Corporation, on behalf of The University of Notre Dame, University of Minnesota and University of Virginia. This paper is based on work funded by NSF Grants 1608834 and 1614320.

\bibliography{report}
\bibliographystyle{spiebib}

\end{document}